\documentclass[letter]{aa}  

\usepackage{graphicx}
\usepackage{txfonts}

\newcommand{\kteff}{$kT^{\infty}_{\mathrm{eff}}$\,}
\newcommand{\nh}{$N_\mathrm{H}$\,}

\newcommand*{\dittoclosing}{---''---}

\usepackage{amstext}

\begin{document}

\title{Unexpected late-time temperature increase observed in the two neutron star crust-cooling sources XTE~J1701$-$462 and EXO~0748$-$676}

\author{A. S. Parikh\inst{1}
 \and
 R. Wijnands\inst{1}
 \and
 J. Homan\inst{2,3}
   \and
 N. Degenaar\inst{1}
 \and
 B. Wolvers\inst{1}
 \and
 L. S. Ootes\inst{1}
 \and
 D. Page\inst{4}
 }

\institute{Anton Pannekoek Institute for Astronomy, University of Amsterdam, Postbus 94249, 1090 GE Amsterdam, The Netherlands\\
\email{a.s.parikh@uva.nl}
\and
SRON, Netherlands Institute for Space Research, Sorbonnelaan 2, 3584 CA Utrecht, The Netherlands
\and
Eureka Scientific, Inc., 2452 Delmer Street, Oakland, CA 94602, USA
\and
Instituto de Astronom\'{i}a, Universidad Nacional Aut\'{o}noma de M\'{e}xico, Mexico D.F. 04510, Mexico
 }

\titlerunning{Unexpected late-time temperature rise in EXO~0748$-$676 and XTE~J1701$-$462}
\authorrunning{A. S. Parikh et al.}

   \date{Received x; accepted x}

  \abstract{Transient low-mass X-ray binaries (LMXBs) that host neutron stars (NSs) provide excellent laboratories for probing the dense matter physics present in NS crusts. During accretion outbursts in LMXBs, exothermic reactions may heat the NS crust, disrupting the crust-core equilibrium. When the outburst ceases, the crust cools to restore thermal equilibrium with the core. Monitoring this cooling evolution allows us to probe the dense matter physics in the crust. Properties of the deeper crustal layers can be probed at later times after the end of the outburst. We report on the unexpected late-time temperature evolution ($\gtrsim$2000 days after the end of their outbursts) of two NSs in LMXBs, XTE~J1701$-$462 and EXO~0748$-$676. Although both these sources exhibited very different outbursts (in terms of duration and the average accretion rate), they exhibit an unusually steep decay of $\sim$7 eV in the observed effective temperature (occurring in a time span of $\sim$700 days) around $\sim$2000 days after the end of their outbursts. Furthermore, they both showed an even more unexpected rise of $\sim$3 eV in temperature (over a time period of $\sim$500-2000 days) after this steep decay. This rise was significant at the 2.4$\sigma$ and 8.5$\sigma$ level for XTE~J1701$-$462 and EXO~0748$-$676, respectively. The physical explanation for such behaviour is unknown and cannot be straightforwardly be explained within the cooling hypothesis. In addition, this observed evolution cannot be well explained by low-level accretion either without invoking many assumptions. We investigate the potential pathways in the theoretical heating and cooling models that could reproduce this unusual behaviour, which so far has been observed in two crust-cooling sources. Such a temperature increase has not been observed in the other NS crust-cooling sources at similarly late times, although it cannot be excluded that this might be a result of the inadequate sampling obtained at such late times. }


   \maketitle

%
\section{Introduction}
Low-mass X-ray binaries (LMXBs) host neutron stars (NS), typically with sub-solar mass companion stars that overflow their Roche lobes. These systems may exhibit outbursts as a result of instabilities in the accretion discs \citep[e.g.][]{lasota2001disc}. The outbursts cause accretion of matter onto the NS, which compresses its crust and results in heat-releasing reactions \citep[e.g.][]{haensel1990non,haensel2008models,steiner2012deep}. These reactions cause the crust to be heated out of thermal equilibrium with the core. They produce $\sim$2 MeV per accreted nucleon in total \citep[e.g.][]{haensel2008models}. Several NS LMXB systems are transient, and the source transitions to quiescence when the outburst ends. During quiescence, the heating reactions are thought to stop, and the crust cools to restore thermal equilibrium with the core. 

Modelling the observed crust-cooling behaviour using theoretical crust-heating and -cooling models \citep[e.g.][]{brown2009mapping,page2016nscool} show that an additional, theoretically unexplained, heat source is present (in the current version of the models) in most NSs in LMXBs. This is known as the `shallow heat source' because it occurs at shallower depths in the crust as compared to the heating processes discussed above. The contribution of the shallow heating is found to be $\sim$1--2 MeV per accreted nucleon in most sources \citep{brown2009mapping,degenaar2011evidence}. However, MAXI J0556$-$332 exhibited $\sim$10--17 MeV per accreted nucleon of shallow heating \citep{homan2014strongly,deibel2015strong,parikh2017different}, but it is currently unknown why this source exhibited such a different amount of shallow heating compared to the other sources.

Although it only makes up $\sim$1\% of the total NS mass by volume, the crust is interesting to study because the density increases by over $\sim$8 orders of magnitude. This is accompanied by changing properties of matter with these increasing densities. Quiescent observations of accretion-heated NSs track the cooling crust and allow us to infer the properties of the dense matter physics present in the crust. These observations, obtained using X-ray telescopes, can only measure the surface temperature of the NS. Observations $\text{of some }$years to several decades after the end of the outburst are needed to probe the deeper layers because the heat from these deeper layers takes this time to be conducted to the surface, where it can be observed \citep{brown2009mapping}. X-ray spectra (in the 0.5--10 keV energy range) of cooling crusts observed in quiescence are characterised by a soft spectral component which may be accompanied by a hard spectral component that is due to an increased presence of photons at energies $\gtrsim$3 keV.

\section{Observations}
We examine the similar late-time quiescent crust cooling behaviour of two systems, XTE~J1701$-$462 and EXO~0748$-$676, that exhibited very different outburst properties. For details of the data analysis method, we refer to Appendix A and \citet{parikh2019consistent}. The details of all the observations we analysed are shown in Table \ref{tab_results}. All errors reported here correspond to 1$\sigma$ errors.

\subsection{XTE~J1701$-$462}
XTE~J1701$-$462 was discovered in January 2006 \citep{remillard2006new} when it exhibited a bright accretion outburst close to its Eddington luminosity. This outburst ended $\sim$1.6 years later in August 2007 \citep{homan2007rossi}, after which the accretion-heated crust was found to cool down. So far, the cooling has been reported up to $\sim$1200 days after the end of its outburst \citep[i.e. t$_\mathrm{q}$, which is also the number of days into quiescence; e.g.][]{fridriksson2010rapid,fridriksson2011variable}. Here, we report on three new observations, extending the cooling baseline to $\sim$3200 t$_\mathrm{q}$.  Several quiescent observations of XTE~J1701$-$462 exhibited enhanced accretion activity in the form of flares \citep[e.g.][]{fridriksson2010rapid}, which are not used for our crust-cooling study. These flaring data are characterised by a large contribution from a hard, power-law-shaped emission component. Therefore, we have only examined non-flaring data here, that is, data for which the flux contribution from the soft spectral component is $>$70 \%.

\subsection{EXO~0748$-$676}
EXO~0748$-$676 exhibited an $\text{about }$24-year-long outburst at an average luminosity corresponding to 5\% of the Eddington luminosity \citep{degenaar2009chandra}. This outburst ended in September 2008 \citep{wolff2008rxte}, and observations made since this time indicate a cooling crust. So far, cooling up to $\sim$1700 t$_\mathrm{q}$ has been reported \citep[e.g.][]{diaz2011xmm,degenaar2009chandra,degenaar2011further,degenaar2014probing,cheng2017cooling}. We report on two new observations, which increases the observed cooling baseline to $\sim$3500 t$_\mathrm{q}$. EXO~0748$-$676 is observed nearly edge-on \citep[at an inclination of $\sim$80$^\circ$;][]{paramar1986discovery} and exhibits eclipses lasting $\sim$8.3 minutes during its $\sim$3.8-hour orbit. We focus on the non-eclipsing `persistent' emission of the source here. Therefore we removed all the data corresponding to the eclipse times, which we determined using the ephemeris reported by \citet{wolff2009eclipse}.

\subsection{Spectral fitting}
\label{sect_spec_fits}
All the XTE~J1701$-$462 and EXO~0748$-$676 spectra were grouped to have a minimum of 25 counts per bin, using the \texttt{grppha} tool. We collectively fit the spectra for the individual sources using \texttt{XSpec}\footnote{https://heasarc.gsfc.nasa.gov/xanadu/xspec/} \citep[version 12.9;][]{arnaud1996xspec} and employing $\chi^2$ statistics.

We fitted the spectra using the NS atmosphere model \texttt{nsatmos} \citep{heinke2006hydrogen}. Our models assumed that the NSs have masses and radii of 1.6 $M_\odot$ and 12 km, respectively. The assumed distances to XTE~J1701$-$462 and EXO~0748$-$676 are 8.8 kpc \citep{lin2009spectral} and 7.4 kpc \citep{galloway2008thermonuclear,degenaar2009chandra}, respectively. For the spectral modelling of both sources, we assumed that the entire NS surface was emitting and hence set the \texttt{nsatmos} normalisation parameter to 1. We used \texttt{tbabs} to model the equivalent hydrogen column density ($N_\mathrm{H}$) with \texttt{WILM} abundances and \texttt{VERN} cross sections \citep{verner1996atomic,wilms2000absorption}. 

The spectra from the two sources could not be well fit (as determined using the reduced $\chi^2$ statistic) with only an \texttt{nsatmos} model because photons at higher energies were present during some observations \citep[at $\gtrsim$3.5 keV; e.g.][]{fridriksson2010rapid, degenaar2011further}. We therefore used an additional power-law component to fit the data. The quality of these data is not good enough  to allow the photon index ($\Gamma$) to vary between the various observations. We therefore tied the $\Gamma$ across the different observations for each source. However, the associated normalisation was allowed to vary. We found that $\Gamma$ was not well constrained (as determined during its error estimation) for several observations. However, the closest fit obtained for $\Gamma$ had rather a low value, and we therefore fixed $\Gamma$ = 1 for both sources.\footnote{We also fitted the data from both sources using $\Gamma$ = 1.5 and 2, and the qualitative trend exhibited by the inferred NS surface temperature did not change.}

\begin{figure}
\centering
\includegraphics[width = \columnwidth]{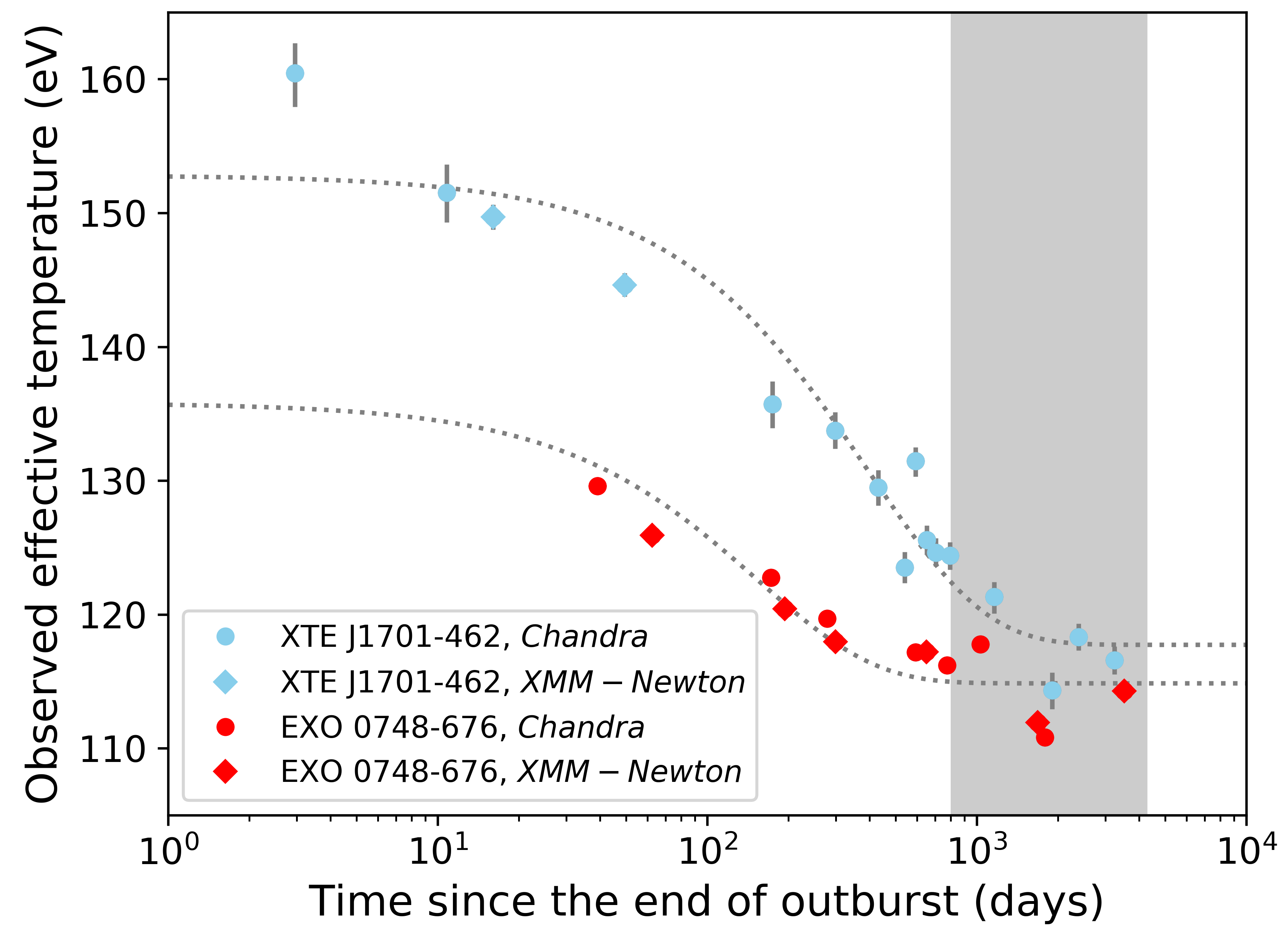}
\caption{The observed effective NS temperature (\kteff) evolution of XTE~J1701$-$462 and EXO~0748$-$676, after the end of the their accretion outbursts, shown in blue and red, respectively. The dotted grey lines are simple functions (exponential decay levelling off to a constant value) used to guide the eye. The solid grey rectangle highlights the region of interest where the decay and subsequent rise in the \kteff is observed in the cooling curves.  The date of the end of outburst for XTE~J1701$-$462 is MJD 54322 and that for EXO~0748$-$676 is MJD 54714 \citep{fridriksson2011variable,degenaar2009chandra}.}
\label{fig_results}
\end{figure}

Initially, the $N_\mathrm{H}$ was left free but was found to be consistent (within the error bars) across all the observations for each source. This $N_\mathrm{H}$ parameter was accordingly tied across the various observations (of each source), but was allowed to vary. The best-fit $N_\mathrm{H}$ determined for XTE~J1701$-$462 and EXO~0748$-$6762 were $(3.14 \pm 0.05) \times 10^{22}$ cm$^{-2}$ and $(4.20 \pm 0.09) \times 10^{20}$  cm$^{-2}$, respectively. The errors on the observed effective NS temperature (\kteff) were determined after fixing the best-fit values corresponding to the normalisation of the power-law component (because the errors on the $\Gamma$ were often unconstrained). The results of our spectral fitting are shown in Table \ref{tab_results} and Figure \ref{fig_results}. The dotted grey lines in Figure \ref{fig_results} are fits of simple exponential decay functions. The figure shows that this function is only a poor fit to the \kteff evolution and is not intended to be a good fit. Its main purpose is to guide the eye to the approximate \kteff evolution of each source.

\begin{sidewaystable*}
\centering
\caption{Log of the observations of XTE~J1701$-$462 and EXO~0748$-$676 obtained after the end of their outbursts and the results of the spectral fitting of these data.$^{\mathrm{a}}$}
\label{tab_results}
\begin{tabular}{lccccccccc}
 \hline
 & Observatory & Observation & MJD & Exposure & \kteff& \texttt{nsatmos} luminosity$^{\mathrm{d}}$ & Total luminosity$^{\mathrm{e}}$ &  \% contribution\tabularnewline
 & & Id$^{\mathrm{b}}$ & & time$^{\mathrm{c}}$ (ksec) & (eV) & ($\times 10^{32}$ erg s$^{-1}$) & ($\times 10^{32}$ erg s$^{-1}$) & of $\Gamma$ \tabularnewline
 \hline
 \multicolumn{10}{c}{{\it XTE~J1701$-$462}} \tabularnewline
1  & {\it Chandra} & 7513 &                     54324.9 & 4.7 &                 160.4$\pm$2.4 &   172.1 $\pm$ 13.9 &   181.6 $\pm$ 13.8       &     5.2    \tabularnewline    
2  & \dittoclosing & 7514 &                     54332.8 & 8.8 &                 151.5$\pm$2.1 &   135.3 $\pm$ 10.4 &   184.9 $\pm$ 11.7       &    26.8    \tabularnewline    
3  & {\it XMM-Newton} & 0413390101 &            54338.0 & 9.5, 20.1, 19.3 &       149.7$\pm$0.9    &128.0 $\pm$  3.9 &   131.3 $\pm$  3.9 &     2.5    \tabularnewline      
4  & \dittoclosing & 0413390201 &               54371.3 & 12.3, 23.2, 23.2 &       144.6$\pm$0.9    &110.9 $\pm$  3.3 &   115.7 $\pm$  3.3 &     4.2    \tabularnewline              
5  & {\it Chandra} & 7515 &                     54496.3 & 19.9 &                135.7$\pm$1.7 &    83.9 $\pm$  5.9 &   120.9 $\pm$  6.0       &    29.8    \tabularnewline    
6  & \dittoclosing & 7516 &                     54620.3 & 27.4 &                133.8$\pm$1.4 &    78.7 $\pm$  4.3 &    88.4 $\pm$  4.4       &    11.0    \tabularnewline    
7  & \dittoclosing & 7517 &                     54753.0 & 39.9 &                129.5$\pm$1.3 &    68.1 $\pm$  3.6 &    91.6 $\pm$  3.6       &    25.7    \tabularnewline    
8  & \dittoclosing & 10063 &                    54862.0 & 50.1 &                123.5$\pm$1.1 &    55.1 $\pm$  2.8 &    59.8 $\pm$  2.8       &     7.9    \tabularnewline    
9  & \dittoclosing &10064, 10889, 10890, 10891& 54914.6 & 53.7 &                131.5$\pm$1.0 &    72.0 $\pm$  3.1 &    86.7 $\pm$  3.0       &    17.0    \tabularnewline    
10 & \dittoclosing & 10065 &                    54974.6 & 62.3 &                125.6$\pm$1.1 &    59.2 $\pm$  2.7 &    71.4 $\pm$  2.6       &    17.0    \tabularnewline    
11 & \dittoclosing & 10066 &                    55027.3 & 65.5 &                124.6$\pm$1.0 &    58.2 $\pm$  2.6 &    69.9 $\pm$  2.5       &    16.7    \tabularnewline    
12 & \dittoclosing & 10067, 12006 &             55117.6 & 59.4 &                124.4$\pm$1.0 &    56.7 $\pm$  2.4 &    63.7 $\pm$  2.4       &    11.0    \tabularnewline    
13 & \dittoclosing & 11087 &                    55480.8 & 56.6 &                121.3$\pm$1.1 &    50.6 $\pm$  2.6 &    56.2 $\pm$  2.6       &     9.9    \tabularnewline    
14 & \dittoclosing & 13730 &                    56227.4 & 64.2 &                114.4$\pm$1.3 &    38.8 $\pm$  2.5 &    53.7 $\pm$  2.3       &    27.8    \tabularnewline    
15 & \dittoclosing & 15800 &                    56706.1 & 113.5 &               118.3$\pm$1.0 &    45.6 $\pm$  1.9 &    61.0 $\pm$  1.8       &    25.3    \tabularnewline    
16 & \dittoclosing & 17870 &                    57561.3 & 98.8 &                116.6$\pm$1.0 &    42.5 $\pm$  2.0 &    48.8 $\pm$  1.9       &    12.9    \tabularnewline
\multicolumn{10}{c}{{\it EXO~0748$-$676}} \tabularnewline
1  & {\it Chandra}      & 9070, 10783 &                 54753.2 & 26.1          &         129.6  $\pm$  0.4       &   67.9   $\pm$   0.9&   79.4  $\pm$  1.0      &       14.5   \tabularnewline    
2  & {\it XMM-Newton}   & 0560180701 &                  54776.4 & 24.2, 29.0, 29.0 &  125.9  $\pm$  0.2       &   60.3   $\pm$   0.4&   64.7  $\pm$  0.5      &        6.8   \tabularnewline    
3  & {\it Chandra}      & 9071, 10871 &                 54886.5 & 23.9          &         122.7  $\pm$  0.4       &   53.5   $\pm$   0.8&   57.4  $\pm$  0.9      &        6.8   \tabularnewline    
4  & {\it XMM-Newton}   & 0605560401 &                  54908.0 & 33.1, 43.2, 42.0 &  120.4  $\pm$  0.2       &   49.6   $\pm$   0.3&   49.6  $\pm$  0.3      &        0.0   \tabularnewline    
5  & {\it Chandra}      & 9072 &                        54992.5 & 26.0          &       119.7  $\pm$  0.4     &   47.8   $\pm$   0.7&   48.0  $\pm$  0.8      &        0.4   \tabularnewline   
6  & {\it XMM-Newton}   & 0605560501 &                  55013.3 & 66.5, 91.4, 92.9  &         118.0  $\pm$  0.1       &   45.3   $\pm$   0.2&   45.3  $\pm$  0.2    &        0.0   \tabularnewline  
7  & {\it Chandra}      & 11059 &                       55306.1 & 25.9          &       117.2  $\pm$  0.5     &   43.6   $\pm$   0.7&   52.9  $\pm$  0.9      &       17.6   \tabularnewline       
8  & {\it XMM-Newton}   & 0651690101 &                  55364.2 & 39.0, 43.0, 40.4 &  117.2  $\pm$  0.2       &   44.0   $\pm$   0.3&   43.9  $\pm$  0.3      &        0.0    \tabularnewline   
9  & {\it Chandra}      & 11060 &                       55489.6 & 26.2          &         116.2  $\pm$  0.5       &   42.2   $\pm$   0.7&   49.1  $\pm$  0.8      &       14.1   \tabularnewline    
10 & \dittoclosing      & 12414 &                       55745.0 & 37.1          &       117.8  $\pm$  0.4     &   44.2   $\pm$   0.6&   51.0  $\pm$  0.7      &       13.3   \tabularnewline       
11 & {\it XMM-Newton}   & 0690330101 &                  56397.2 & 81.5, 99.1, 99.2 &  111.9  $\pm$  0.1       &   35.9   $\pm$   0.2&   35.9  $\pm$  0.2      &        0.0   \tabularnewline    
12 & {\it Chandra}      & 14663 &                       56505.1 & 41.1          &         110.8  $\pm$  0.4       &   33.9   $\pm$   0.6&   35.8  $\pm$  0.6      &        5.3   \tabularnewline    
13 & {\it XMM-Newton}   & 0824420101 &                  58240.3 & 63.2, 74.1, 74.1 &  114.3  $\pm$  0.1       &   39.0   $\pm$   0.2&   40.9  $\pm$  0.2      &        4.7   \tabularnewline     
\hline
 \multicolumn{10}{p{22cm}}{$^{\mathrm{a}}$All errors reported are 1$\sigma$ errors.} \tabularnewline
  \multicolumn{10}{p{22cm}}{$^{\mathrm{b}}$ If multiple observations Ids are listed,  the spectra obtained from these observations were combined into one single spectrum, resulting in a single temperature estimate.} \tabularnewline
 \multicolumn{10}{p{22cm}}{$^{\mathrm{c}}$The exposure times are reported after the removal of background flares and eclipses. The {\it XMM-Newton} exposure times correspond to data from the pn, MOS1, and MOS2 times, respectively.} \tabularnewline
 \multicolumn{10}{p{22cm}}{$^{\mathrm{d}}$The \texttt{nsatmos} luminosity refers to the total unabsorbed luminosity corresponding only to the contribution from the \texttt{nsatmos} component for the 0.5--10 keV energy range.} \tabularnewline
\multicolumn{10}{p{22cm}}{$^{\mathrm{e}}$The total luminosity refers to the total unabsorbed luminosity (corresponding to the contributions from the \texttt{nsatmos} and power-law components) for the 0.5--10 keV energy range.} \tabularnewline
\end{tabular}
\end{sidewaystable*}

\section{Results and discussion}
\subsection{Results}
We present the observed quiescent temperature evolution of two NS crust-cooling sources, XTE~J1701$-$462 and EXO~0748$-$676, in Figure \ref{fig_results} and Table \ref{tab_results}. Both sources exhibit a cooling evolution similar to those observed for other sources up to $\sim$1000 t$_\mathrm{q}$. However, observations obtained at $\sim$2000 t$_\mathrm{q}$ show a steeper cooling decay in both sources than their previous observed evolution \citep[as is predicted by some of the models for XTE~J1701$-$462, presented in][]{page2013forecasting}. During this stage, the sources decayed by $\sim$6--7 eV over $\sim$650--750 days. However, unanticipated for sources that exhibit crust {\it cooling}, the observations of both sources (at $>$2000 t$_\mathrm{q}$) exhibit a {\it rise} in the observed \kteff (by $\sim$3 eV over $\sim$500--2000 days) instead of a further decay. The observed rise for XTE~J1701$-$462 and EXO~0748$-$676 was significant at the 2.4$\sigma$ and 8.5$\sigma$ level, respectively. Extraordinarily, the magnitude of the observed rise in \kteff and the time after the end of the outburst at which both sources exhibited this rise are similar\footnote{The exact times at which the sources were observed to decay or exhibit their subsequent rise are better constrained for XTE~J1701$-$462 than for EXO~0748$-$676 due to the observational sampling.} although they exhibited different outburst properties. Other cooling sources observed at late times do not exhibit a similar behaviour \citep[see Figure 2 of][for the late-time crust-cooling curve of several sources]{wijnands2017review}. However, this may be a result of the sparse observational sampling obtained, therefore it remains possible that these sources behaved in a similar fashion.

The solid grey rectangle in Figure \ref{fig_results} highlights the area of interest in the cooling curves. A decrease and subsequent increase in the \kteff is observed there.

\subsection{Examining the spectra and their fits}
A potential explanation for the sudden increase in \kteff at late times in both sources might be that the sources exhibited low-level accretion events during these observations. Such events may affect the spectra so as to result in an inferred increase in \kteff (likely without an actual physical increase in the crust temperature). However, such events tend to be characterised by an increased contribution from a hard spectral component \citep[contributing $\sim$50\% in the 0.5--10 keV energy range; e.g.][]{fridriksson2010rapid,chakrabarty2014hard,wijnands2015low,d2015radiative}. The last few observations for both our sources during which a rise in the \kteff was observed do not exhibit a dominant contribution from the power-law component (the contribution was $<<$50\% in the 0.5--10 keV energy range, see Table \ref{tab_results}), suggesting that the observed \kteff rise is not a result of accretion events.

A change in the actual \nh compared to its assumed value might also affect the spectral fit parameters and result in an observed increase in the \kteff \citep[as was discussed for the crust-cooling source MXB 1659$-$29;][]{cackett2013change}. This is especially pertinent for EXO~0748$-$676 because the source is viewed at a high inclination, and the build-up of the disc in quiescence might increase the \nh towards the source \citep[as was suggested to occur in MXB 1659$-$29 as well;][]{cackett2013change}. However, the observed increase in \kteff is not a result of our \nh assumptions because the \nh remains consistent within the error bars (when left free to vary during our spectral fits; see Section \ref{sect_spec_fits}) between the various spectra of EXO~0748$-$676 (and for XTE~J1701$-$462 as well, for which we did a similar test).

\subsection{Might it still be crust cooling?}
The late-time \kteff evolutions of XTE~J1701$-$462 and EXO~0748$-$676 are very similar, exhibiting a steep decay followed by a rise in the observed \kteff after the end of their accretion outbursts. This suggests that the cause of the observed behaviour may have the same physical explanation.  Below we discuss possible reasons for crust-cooling sources to exhibit this unexpected late-time rise in the observed \kteff. 

\subsubsection{Investigating the possibility of an intermediate outburst}
An accretion outburst occurring at an intermediate time (i.e. after the end of the observed main outburst and before the observed \kteff increase) might result in reheating of a cooling crust and might explain (if such outburst indeed occurred) the observed increase in temperature \citep[as was observed for MAXI J0556-332;][] {parikh2017different}. An accretion outburst might also result in a change in the composition of the envelope (the outer $\sim$100 m of the crust that translates the temperature at the top layers of the crust to the effective surface temperature), thereby changing its thermal conductivity and the observed \kteff \citep{brown2002variability}. It may also be a combination of both these effects. 

The observed decay and subsequent rise  in the \kteff (with respect to our limited observational sampling) is separated by $\sim$1.3 and $\sim$4.8 years in XTE~J1701$-$462 and EXO~0748$-$676, respectively. We investigated the archival data from both these sources to examine whether they exhibited any intermediate outbursts before the observed rise in temperature. We used data from the Burst Alert Telescope (BAT) on board the {\it Neil Gehrels Swift Observatory} and the Monitor of All-sky X-ray Image (MAXI) on board the {\it International Space Station}. However, no evidence for enhanced activity was exhibited by either source. Thus, if these sources experienced any intermediate outburst, it would have been rather faint, having a maximum luminosity of $\sim5 \times 10^{35}$ erg s$^{-1}$ \citep[][]{hiroi2009maxi,krimm2013swift}, such that it is not detected by these all-sky monitors. However, we note that XTE~J1701$-$462 typically cannot be observed for $\text{about three}$ months of the year (from early November to the end of January) because of the constraint imposed by the location of the Sun in the sky. It might have exhibited a brighter outburst during this time. EXO~0748$-$676 is never Sun constrained and can be observed throughout the year. Short and faint accretion episodes are not expected to significantly heat up the crust \citep[e.g.][]{wijnands2013low}.

\subsubsection{Ocean freezing}
The NS crust may melt to form an `ocean' as a result of the heat released during the accretion outburst. During quiescence, this ocean cools and solidifies \citep[e.g.][]{medin2011compositionally}. Heavy elements preferentially freeze out from the ocean, enlarging the solid crust and setting up convection in the ocean \citep[e.g.][]{medin2011compositionally,medin2014signature,medin2015time,mckinven2016survey}. This affects the evolution of the modelled crust-cooling curve as compared to the scenario during which no chemical convection is accounted for. \citet{medin2014signature} presented some crust-cooling studies that included the solidification of the ocean. Their Figure 1 shows that a decrease followed by a rise at late times can be observed under certain assumptions \citep[see][for more details]{medin2014signature,medin2015time}. This is reminiscent of what we have observed for our sources.

\subsubsection{Nuclear pasta}
The late-time evolution of crust cooling NSs, as studied here, provide the exciting possibility to probe the physics in the deeper crust where nuclear pasta may be present \citep[e.g.][]{caplan2017astromaterial}. Nuclear pasta refers to the disordered, non-spherical matter, expected at the large densities near the crust-core interface, as the short range nuclear attraction competes with the long range Coulomb repulsion \citep[e.g.][]{pethick1998liquid}. This disordered nuclear pasta is expected to be characterised by low thermal conductivity, forming a heat barrier \citep[e.g.][]{horowitz2015disordered}. 

During accretion, the heat-producing reactions result in the overall heating of the crust. If the NS contained low thermal conductivity pasta and a relatively cold core, the heat would flow inwards during accretion and not contribute much to the observed crust-cooling evolution in quiescence due to the thermal profile in the crust. Alternatively, if the core is relatively hot, the crustal thermal profile would be such that the heat flux during quiescence, from the core outwards towards the surface would not change much. We therefore hypothesise that the presence of pasta (in an NS with either a cold or a hot core) would not explain the decay and subsequent rise in our observed \kteff evolution because this would only produce a decay in the cooling curve and no subsequent rise. Cooling-curve modelling including a pasta layer has previously been carried out for several crust-cooling sources \citep[MXB 1659$-$29, MAXI J0556$-$332, and Terzan 5 X-3;][]{horowitz2015disordered,parikh2017different,ootes2020} which showed the expected behaviour we hypothesised for XTE~J1701$-$462 and EXO~0748$-$676 here.

\subsubsection{Time-variable properties of a deep crust layer}
Although nuclear pasta cannot reproduce the decay in the observed \kteff, the decay may be reproduced by a relatively deep crust layer that exhibits a time-variable thermal conductivity that suddenly increases and leads to the observed decrease in \kteff, and then subsequently decreases again and leads to a sudden rise in the \kteff. However, there does not appear to be any physical motivation for a such a time-dependent thermal conductivity. 

\vspace{0.8cm}

\noindent Our study of the crust-cooling sources, XTE~J1701$-$462 and EXO~0748$-$676, has shown an unexpected rise in the temperature at late times in quiescence. The physical explanation for this behaviour is unknown. Intensive theoretical investigations (beyond the scope of this paper) may provide an insight into the underlying physics and investigate whether the new physical explanations invoked by the theoretical models would only affect the late-time crust-cooling evolution or have a broader effect on the physics inferred across all timescales. 

\vspace{0.5cm}
\begin{acknowledgements}
This work benefitted from support by the National Science Foundation under Grant No. PHY-1430152 (JINA Center for the Evolution of the Elements). AP, RW, and LO were supported by a NWO Top Grant, Module 1, awarded to RW. AP and ND are supported by an NWO/Vidi grant awarded to ND. JH acknowledges financial support that was provided by the National Aeronautics and Space Administration through Chandra Award Numbers GO2-13060X, GO4-15048X, and GO6-17048X issued by the Chandra X-ray Center, which is operated by the Smithsonian Astrophysical Observatory for and on behalf of the National Aeronautics Space Administration under contract NAS8-03060. DP is partially supported by the Consejo Nacional de Ciencia y Tecnolog{\'\i}a with a CB-2014-1 grant $\#$240512. 
\end{acknowledgements}

\bibliographystyle{aa}


\begin{appendix}
\label{app}
\section{Data analysis}
We processed the quiescent observations from both sources using the same software version. We briefly describe the processes below. 

\subsection{Chandra}
The {\it Chandra} data were processed using CIAO\footnote{https://cxc.harvard.edu/ciao/} (v4.11). The data were examined for background flares by making light curves of the background (excluding all bright sources), but no evidence of any strong flaring was detected. A source region centred on the source location with a radius of 3 arcsec was used to extract the spectra for both sources. The background region we used for the spectral extractions of all observations was annular, with an inner and outer radius of 10 and 20 arcsec, respectively. The annulus was centred on the source location. The \texttt{specextract} tool was used to generate the spectra from the various observations, along with the aperture-corrected auxiliary response files and the redistribution matrix files. We combined {\it Chandra} spectra that were close in time (i.e. obtained $\lesssim$3 days apart) to obtain data with higher quality. This was done using the \texttt{combine{\_}spectra} tool. The observations combined into single spectral intervals are shown in Table \ref{tab_results}.

\subsection{XMM-Newton}
The {\it XMM-Newton} data were processed using SAS\footnote{https://www.cosmos.esa.int/web/xmm-newton/download-and-install-sas} (v18.0) We inspected all the {\it XMM-Newton} data for any evidence of background flares by examining the 10--12 keV and $>$10 keV data for the pn and MOS detectors, respectively. All occurrences of background flaring were removed by discarding data from time intervals when flaring was observed. On average, we discarded $\sim$20 \% of the data as a result of such flaring.

Source regions with a radius of 25 arcsec, placed on the source (as suggested by the \texttt{eregionanalyse} tool), were used to extract the spectra of both sources. A circular background region with a radius of 50 arcsec whose location was recommended by the \texttt{ebkgreg} tool was used for the spectral extraction of all observations we examined here. The ancillary response function and response matrix files were generated using \texttt{arfgen} and \texttt{rmfgen}.

\end{appendix}

\end{document}